# Spectrum Pooling in MmWave Networks: Opportunities, Challenges, and Enablers

Federico Boccardi, Hossein Shokri-Ghadikolaei, Gabor Fodor, Elza Erkip, Carlo Fischione,
Marios Kountouris, Petar Popovski, and Michele Zorzi

## 1 Abstract

Motivated by the intrinsic characteristics of mmWave technologies, we discuss the possibility of an authorization regime that allows spectrum sharing between multiple operators, also referred to as *spectrum pooling*. In particular, considering user rate as the performance measure, we assess the benefit of coordination among the networks of different operators, study the impact of beamforming both at the base stations and at the user terminals, and analyze the pooling performance at different frequency carriers. We also discuss the enabling spectrum mechanisms, architectures, and protocols required to make spectrum pooling work in real networks. Our initial results show that, from a technical perspective, spectrum pooling at mmWave has the potential for a more efficient spectrum use than a traditional exclusive spectrum allocation to a single operator. However, further studies are needed in order to reach a thorough understanding of this matter, and we hope that this paper will help stimulate further research in this area.

F. Boccardi's work was carried out in his personal capacity and the views expressed here are his own and do not reflect his employer's ones. (e-mail: federico.boccardi@ieee.org).

H. Shokri-Ghadikolaei and C. Fischione are with School of Electrical Engineering, KTH Royal Institute of Technology, Stockholm, Sweden (email: hshokri, carlofi@kth.se).

E. Erkip is with the NYU WIRELESS Center, Polytechnic School of Engineering, New York University, Brooklyn, NY, USA (e-mail: elza@poly.edu).

M. Kountouris is with the Mathematical and Algorithmic Sciences Lab, France Research Center, Huawei Technologies Co. Ltd. The views expressed here are his own and do not represent Huawei's ones. (email: marios.kountouris@huawei.com).

P. Popovski is with the Department of Electronic Systems, Aalborg University, Aalborg, Denmark (e-mail: petarp@es.aau.dk).

M. Zorzi is with the Department of Information Engineering, University of Padova, Padova, Italy (e-mail: zorzi@dei.unipd.it).

G. Fodor was funded by Wireless@KTH and his work is part of the Wireless@KTH project BUSE. The work of E. Erkip was partially supported by the National Science Foundation under Grant 1547332. The work of M. Zorzi was partially supported by NYU-Wireless. The work of P. Popovski was partially supported by the ERC Consolidator Grant, Horizon 2020.

## 2  Introduction

The demand for mobile wireless services is predicted to increase significantly in the next years. The scarcity of available microwave spectrum, which cannot satisfy this increased demand, has led to the emergence of mmWave as the new frontier of wireless communication. Recently, as part of the harmonization process that will lead to a new mobile spectrum, the 2015 World Radio Conference selected different bands, ranging from about 24 GHz to 86 GHz, for further studies for use in future 5G systems. Unfortunately, the availability of spectrum for mobile services presents limitations even at mmWave frequencies, particularly if one considers the requirements of other systems that may also use these bands in the future, including satellite, and fixed services. This is further exacerbated if we also consider the need to license mobile bands to multiple operators and thereby foster healthy competition in the market. Therefore, it is essential to seek an optimal use of the spectrum, with the ultimate goal of maximizing the benefits for citizens.

The type of spectrum access scheme plays a fundamental role in achieving an efficient usage of the spectrum. Spectrum sharing allows multiple service providers to access the same band for the same or different uses. This paper investigates the case of spectrum sharing for the same use – mobile services – between different mobile operators, also referred to as *spectrum pooling*. The specific features of mmWave frequencies, for example, the propagation characteristics and the operation based on directional beamforming, are expected to be critical enablers for spectrum pooling, but also call for judiciously designed new paradigms.

### 2.1  Background on spectrum pooling

Spectrum pooling has been recently considered for cellular systems at microwave frequencies. For example, in [1] (and the references therein), it was shown that orthogonal spectrum pooling, whereby frequency channels are dynamically but exclusively allocated to one operator at a time, results in significant throughput gains, in the order of 50–100%. In addition, if frequency channels can be allocated simultaneously to multiple operators, called non-orthogonal spectrum pooling, further gains can be obtained. To achieve these gains, coordination mechanisms, both within an operator, hereafter called intra-operator coordination, and among different operators' networks, hereafter called inter-operator coordination may be required.

There are two different architectural approaches to spectrum sharing; namely, with or without sharing of radio access network (RAN) infrastructure. The benefits of spectrum pooling with RAN sharing are discussed in [2]. In the context of microwave HetNets, it was shown in [3] that a RAN sharing strategy might be optimal for small cells. When spectrum pooling is used without RAN sharing, interference becomes the main limiting factor, and simple interference management techniques may lead to suboptimal spectrum utilization. Therefore, interference-aware techniques have been studied, and the benefits of spectrum pooling with smart scheduling have been discussed in, for example, [4].

Recent works have also considered the benefits of spectrum sharing at mmWave frequencies. In [5], a mechanism that allows two different IEEE 802.11ad access points to transmit over the same time/frequency resources was proposed. This is realized by introducing a new signaling report broadcasted by each access point, in such a way as to establish an interference database to support scheduling decisions. A similar approach was proposed in [6] for mmWave cellular systems, with both centralized and distributed inter-operator coordination. In the centralized case, a new architectural entity receives information about the interference measured by each network and determines which links cannot be scheduled simultaneously. In the decentralized case, the victim network sends a message to the interfering network with a proposed coordination pattern. The two networks can further refine the coordination pattern via multiple stages. Reference [7] studied the feasibility of spectrum pooling in mmWave cellular networks, and shows that under certain conditions (for example, ideal antenna pattern), spectrum pooling may be beneficial, even without any coordination between different operators.

## 2.2  Contributions of this work

The aim of this paper is twofold.

First, we aim to present technical evidence, by means of simulation results, which reveals under which assumptions and conditions spectrum pooling at mmWave frequencies is beneficial. To this end, starting from the studies in [6] and [7], we provide substantial extensions to them. In contrast to [6], where the emphasis is on the multiple access control (MAC) layer, we jointly consider physical and MAC layers. The emphasis in [7] is on the physical layer without coordination, but here we consider the effects of both intra- and inter-operator coordination. We show that, while coordination may not be needed under ideal assumptions, it does provide substantial gains when considering realistic channel and interference models and antenna patterns. Moreover, we evaluate the impact of beamforming, antenna array size, different carrier frequencies, and different BS densities on the pooling performance.

Second, using the insights derived from our quantitative analysis, we discuss the technical enablers required to make spectrum pooling work under realistic assumptions and constraints. We discuss the tradeoffs among the architectural solutions, the type of coordination, the amount and type of information exchange required, and the new enabling functionalities. We argue that further works are needed to assess other spectrum access regimes, for example, those built on the aggregation between licensed and license-exempt spectrum.

We believe that this work makes an important contribution towards answering the following fundamental question related to future mmWave networks: For a given amount of spectrum for mobile applications at a given mmWave frequency, what is the access scheme that allows its optimal utilization?

This work is particularly timely given discussion on this topic has just started in 3GPP and ITU-R, and it must be supported by rigorous studies and solid technical evidence on spectrum pooling performance and on the related technology enablers.

The remainder of this paper is organized as follows. Section 3 provides an assessment of the benefits of spectrum pooling via a technical analysis. Inspired by these technical results, we discuss the protocol and architectural enablers for spectrum pooling in Section 4. We highlight future research directions in Section 5, followed by concluding remarks given in Section 6.

## 3   Performance assessment

In this section, we discuss our initial assessments of spectrum pooling performance at mmWave, in terms of UE rate enhancement. The analysis is based on ideal assumptions and is aimed at unveiling the potential of spectrum pooling, rather than quantifying the gains in a realistic setup.

As a starting point, we note that with spectrum pooling each user equipment (UE) has access to a larger bandwidth, at the expense of a potentially lower signal-to-interference-plus-noise ratio (SINR) due to the increased inter-operator interference. In addition, a larger bandwidth leads to a correspondingly increased noise power. Inter-operator interference can be tackled in two complementary ways: either more directional beams or inter-operator coordination. Under the assumption of a constant array size, the use of higher frequencies allows deploying more antenna elements per array at both base stations (BSs) and UEs, hence increasing the beamforming gain. The gains of inter-operator coordination are more complex and are a function of the specific implementation and of the supporting architecture. In this section, we consider a centralized implementation based on joint beamforming and user association. In Section 4, we will then discuss the impact of more practical coordination schemes.

### 3.1   Simulation Scenarios

We consider a mmWave system where antenna and channel models are as in [8], and BSs and UEs are randomly distributed as in [7]. Without loss of generality, we assume analog beamforming both at the BSs and at the UEs, four operators, and a total bandwidth of 1200 MHz. We consider three spectrum pooling scenarios:

- Exclusive: each operator uses a 300 MHz exclusive bandwidth;
- Partial pooling: operators 1 and 2 share the first 600 MHz, and Operators 3 and 4 share the second 600 MHz;
- Full pooling: all operators share the whole 1200 MHz bandwidth.

We consider two coordination scenarios:

- Baseline, w/o inter-operator coordination: only BSs belonging to the same operator coordinate to perform a joint user association and beamforming.
- Inter-operator coordination: coordination is extended to BSs belonging to different operators.

We note that we use an ideal coordination scheme, with the aim of providing a performance upper bound. In particular, we consider a centralized coordination approach where a central entity jointly selects users and calculates beams so as to maximize the user rates based on a proportionally fair criterion. The central entity is assumed to have perfect knowledge of the long-term channel parameters for each user and each BS in the network, and of the load of each BS. We assume that different BSs are synchronized, and that there is no delay in the interface between the BSs and the central entity.

### 3.2 Simulation Results

Fig. 1 illustrates the gain of partial and full pooling w.r.t. an exclusive spectrum allocation, under the assumption of no inter-operator coordination. We show the $5^{th}$, $50^{th}$, and $95^{th}$ percentiles of the UE downlink rates at 32 GHz, assuming a BS density of 100 BSs/km$^2$ and a user density of 800 UEs/km$^2$. These results also apply to 28 GHz.

In Fig. 1(a), we assume a 32x32 uniform planar array (UPA) at each BS and a 4x4 UPA at each UE. Moreover, without loss of generality, each BS uses six RF chains, such that it can create simultaneously up to six analog beams. We observe in Fig. 1(a) that most UEs benefit from spectrum pooling. In particular, both partial and full pooling enhance the $5^{th}$, $50^{th}$, and $95^{th}$ percentiles compared to the baseline (i.e., exclusive). In Fig. 1(b), we repeat the previous comparison under the assumption of a single omnidirectional antenna at the UE, as a way to study the effect of a less effective beamforming. In this case, partial and full pooling lead to worse performance for the $5^{th}$ percentile UEs and (for the case of full pooling) for the $50^{th}$ percentile UEs, due to an increased inter-operator interference.

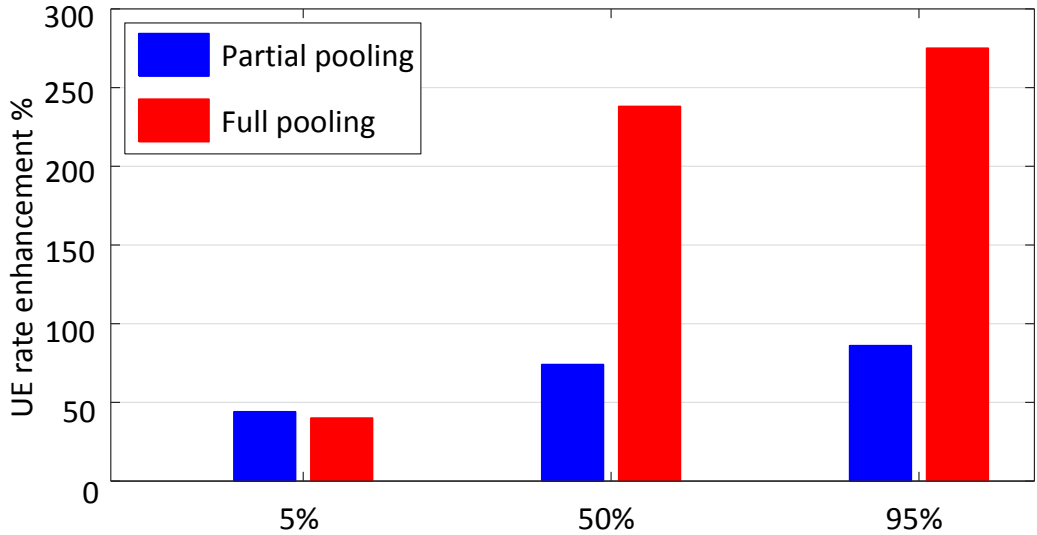

a) 4x4 UPA at the UE.

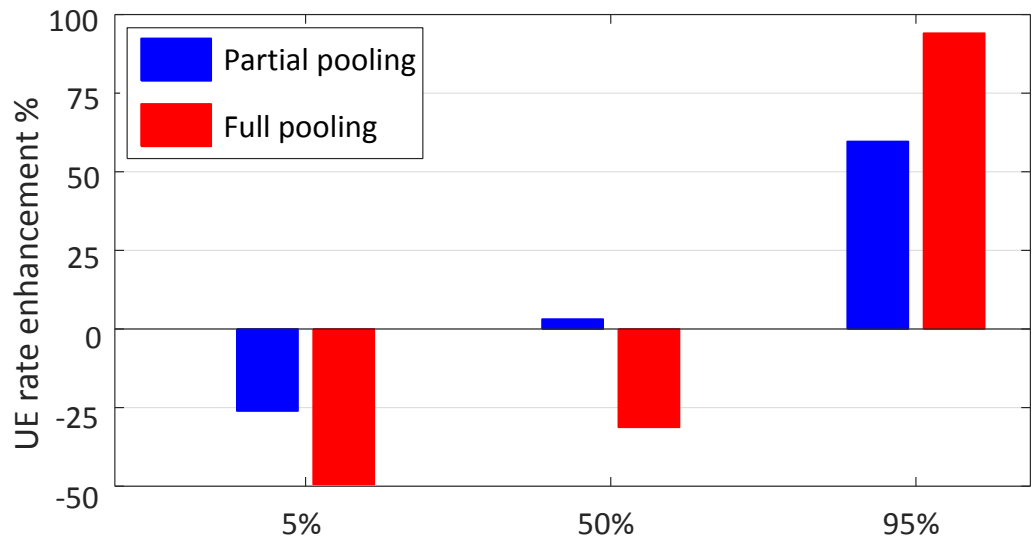

b) Omnidirectional antenna at the UE.

Fig. 1. Pooling performance at 32 GHz, under the assumption of no inter-operator coordination. The baseline is a system with exclusive spectrum allocation.

Fig. 2 shows the impact of the operating frequency and the BS density on the performance. We consider transmissions at 32 GHz and 73 GHz. We keep the dimension of the antenna array constant as a function of the frequency, i.e., at 73 GHz we consider twice the antenna elements in each dimension w.r.t. 32 GHz, at both BS and UE. We plot the gain of full pooling compared to an exclusive spectrum allocation, for different values of the BS density: 50 and 200 BSs/km$^2$

(corresponding to a cell radius of 80m and 39 m, respectively). We note that, without inter-operator coordination, increasing the BS density of individual operators exacerbates the inter-operator interference and reduces the benefits of spectrum pooling. For example, at 32 GHz when going from 50 BSs/km$^2$ to 200 BSs/km$^2$, spectrum pooling at the 5$^{th}$ percentile users is reduced from 50% to almost 0%. This effect is less pronounced at 73 GHz, due to the higher directionality of the beams. Fig. 2 shows that inter-operator coordination is very effective in very dense deployments (200 BS/km$^2$) and for the weakest UEs (5$^{th}$ percentile UEs). Moreover, full coordination is more critical at 32 GHz than at 73 GHz, due to the fact that beamforming by itself is not sufficient to protect the weakest users from inter-network interference.

### 3.3 Discussion

The results presented above clearly indicate that, under ideal assumptions, spectrum pooling is beneficial, and it provides gain at the 5$^{th}$, 50$^{th}$ and 95$^{th}$ user percentiles. However, more work is required to assess the impact of real-world effects. Our analysis indicates that beam directionality is a critical enabler, which is consistent with the results in [7]. For example, when using a single omnidirectional antenna at the UE, pooling performance drastically decreases. We expect beamforming to be even less effective under more realistic assumptions. As a matter of fact, beamforming gains may be significantly affected by real-world effects like pilot contamination, imperfect channel estimation and mobility. BS density also has an impact on the performance: in very dense deployments, even under ideal assumptions about beamforming, the weakest users suffer from inter-operator interference. Differently from [7], we found that inter-operator coordination is required, especially for the weakest users. Moreover, our results show that inter-operator coordination is more critical at 32 GHz than at 73 GHz. We note that, while in this study we consider a centralized coordination scheme, the impact of more realistic (for instance, distributed) coordination approaches should be further studied. This indicates that another critical enabler is a reliable control channel for exchanging coordinating information [9]. We will further discuss this and other related aspects in the following section.

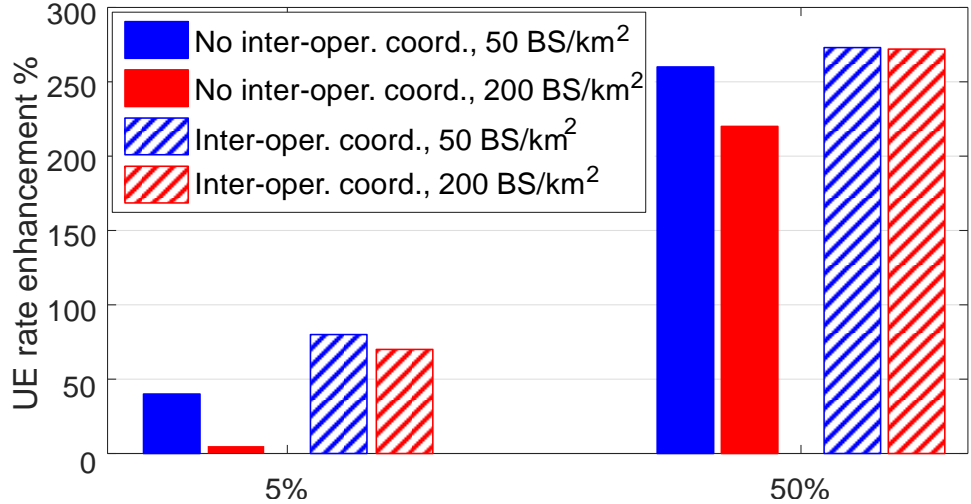

a) 32 GHz, 4x4 UPA at the UE, 32x32 UPA at the BS.

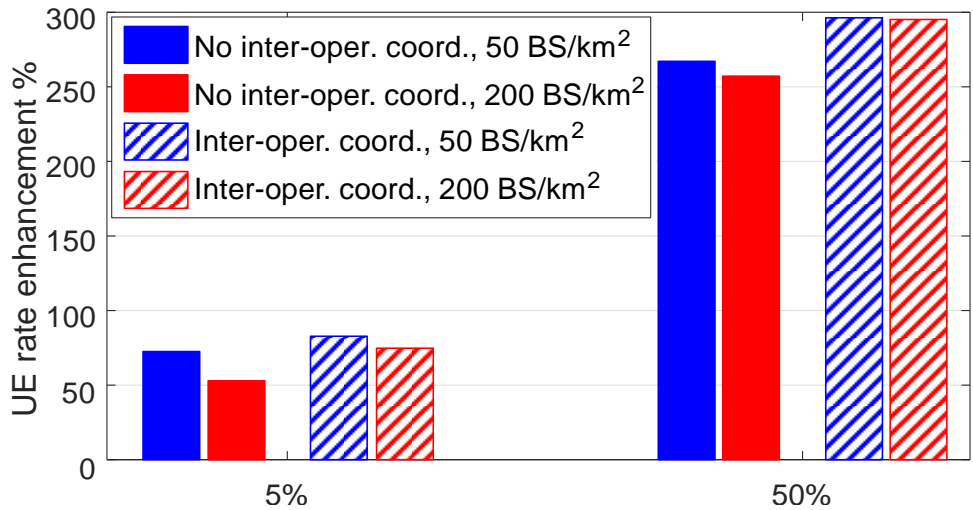

b) 73 GHz, 8x8 UPA at the UE, 64x64 at the BS.

Fig. 2. Full pooling performance, w/o and with inter-operator coordination.

## 4 Protocol and architectural enablers

Our assessments in Section 3 showed that coordination between different network operators is beneficial. However, in Section 3 we considered an ideal coordination framework by means of a logical central entity. In the following we further examine this aspect and discuss different coordination types, different supporting architectures and different functional enablers.

### 4.1 Types of spectrum pooling coordination

Coordination is distributed when decisions are made in each operator domain, aided by the exchange of supporting information. Such supporting information can be exchanged prior to resource allocation decisions. For example, a participating operator may report the set of subbands, resource blocks or beams within the shared spectrum pool that are not used within a geographical area or in a given cell, for example, due to low traffic load. In addition, information between participating operators may be exchanged in a reactive fashion. For instance, high interference levels measured in subbands within the shared spectrum pool can trigger a request from a participating operator to its peer operator to reschedule some of the served traffic to other resources. Such inter-operator distributed coordination schemes can be seen as extensions of the RAN sharing scenarios studied by the 3GPP in TR 22.852 [10]. With centralized coordination (like the one we used in Section 3), the actions are decided by a logical central entity, such as a spectrum broker [11] or a module for making network policy, supported by a network-wide database [12]. We note that the feasibility of centralized coordination also depends on the latency of the exchanged information. If the latency is sufficiently low, then any architecture that supports distributed coordination can support centralized coordination as well, by electing one of the distributed entities as a leader and thus a logical coordination center. We will further discuss the architectural aspect in the next section.

With an uncoordinated approach, network operators do not exchange any information and make independent decisions on how to allocate spectrum. However, common rules need to be in place in order to ensure equitable spectrum allocation.

Inter-operator coordination can be near real-time (operating on a time scale of hundreds of ms) or long-term (operating on the time scale of seconds, minutes, or even coarser scales). In the first case, BSs coordinate at the level of the resource block scheduling. Such near real-time coordination is similar to the X2 application protocol based intra-operator mechanisms, which can be deployed in LTE networks, including inter-BS signaling schemes on traffic load and high interference indications [13]. In the second case, spectrum usage is supported by information exchange on a coarse time scale, typically implemented as part of the operation and maintenance (O&M) infrastructure in each participating operator's network. This type of long-term spectrum usage coordination can operate on the basis of an inter-operator agreement on a usage portion of the pooled spectrum resources during different times of the day or associated with specific days of the week. It can also include a maximum level of energy emissions on specific parts of the spectrum pool. Long-term spectrum usage coordination can be conveniently realized by information exchange at the O&M level rather than employing protocol messages between RAN nodes, as in the case of near-real-time coordination schemes.

## 4.2 Supporting architectures

The different spectrum pooling mechanisms discussed above can be implemented through different supporting architectures. Fig 3 gives a high level summary of the main alternatives as follows:

- *Interface at the RAN:* Alternative (a) in Fig 3 refers to the introduction of a new interface (or an extension of the X2 interface used for LTE [13]) between BSs belonging to different networks, to enable distributed coordination. From a logical architectural perspective, Alternative (a) allows a fast information exchange between two different networks,[1] and therefore near real-time spectrum pooling is possible.

- *Interface at the core network (CN):* Alternative (b) refers to an architecture where the interface between the different networks is at the CN. Due to the latency involved, Alternative (b) does not enable real-time spectrum pooling. On the other hand, CN level coordination can impact a large number of cellular BSs by exchanging a few protocol messages, since typically a large number of BSs are associated with a few CN nodes.

- *RAN sharing:* Alternative (c) refers to an architecture where two or more network operators share the BSs. In other words, a single baseband unit serves users associated to the different network operators in the sharing agreement. As resource allocation and scheduling decisions are made by a single unit, Alternative (c) is an effective way to implement real-time centralized coordination.

- *CN sharing:* Alternative (d) refers to an architecture where two or more network operators share the CN. In the same way as Alternative (c), Alternative (d) allows centralized coordination. However, in this case real-time spectrum pooling is not possible for the same reasons discussed for Alternative (b).

- *Spectrum broker:* Alternative (e) refers to an architecture where coordination is implemented by means of a spectrum broker. A spectrum broker is a central resource management entity that grants spectrum resources on an exclusive basis during some time window [11].

---

[1] We note that the real-world latency depends on the specific characteristics of the wired/wireless transport network technology and topology. For example, ultra-dense mmWave networks will likely rely on in-band wireless transport possibly involving multi-hop routes that may result in increased latency between wireless access points and core network nodes.

– *Uncoordinated spectrum pooling:* Alternative (f) refers to the case where the network operators do not coordinate. When the number of networks in the pool is not limited, uncoordinated spectrum pooling is reminiscent of a license-exempt regime. For example, in wireless LANs (for example, Wi-Fi) real-time spectrum pooling is realized through uncoordinated operation.

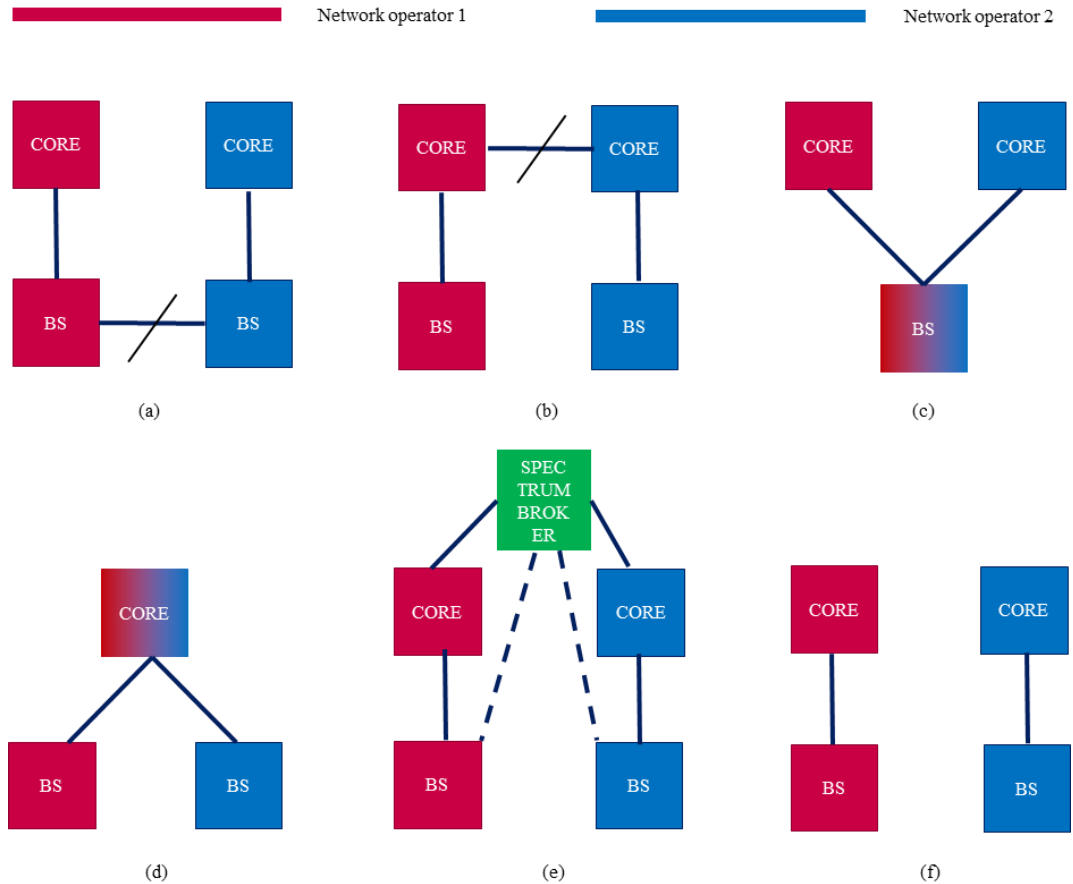

Fig 3. Architectural solutions supporting spectrum pooling between two different network operators. (a): interface at the RAN (base station), (b): interface at the CN, (c): RAN sharing, (d): CN sharing, (e): via a spectrum broker, (f): uncoordinated.

### 4.3 Supporting functions

Spectrum pooling mechanisms require different supporting functions depending on the type of coordination and on the architectural solution. In the following we discuss the most prominent supporting functions.

- Spectrum sensing: Spectrum sensing and dynamic frequency/channel selection (DFS/DCS) are solutions in which systems participating in spectrum pooling dynamically select their operating frequency range based on measurement results [11], [12]. These measurements can be overall energy or reference signal detection. DFS/DCS are typically not considered as a reliable method due to the well-known hidden node problem. However, spectrum sensing and DFS/DCS can have a supporting role, for example, to identify spectrum subbands that have the least instantaneous traffic load so that sharing overhead is minimized. Spectrum sensing may be supported by new radio interface capabilities such as the capability at the UE of sensing interference originated by a non-serving network that participates in the spectrum pool. We note that the lower the level of coordination between different network operators, the more important the role of spectrum sensing and DFS/DCS.

- Enhanced channel state information (CSI) acquisition and exchange techniques that help learn the channel at both transmitters and receivers. Given the importance of accurate narrow beamforming as an enabler of spectrum pooling and the sensitivity of precoding to reference signal contamination, participating networks may create clean subbands used for training signals. For example, networks operating in time division duplexing (TDD) mode and relying on channel reciprocity to acquire CSI both for UL reception and DL precoding may reserve own (not shared) pilot sequences that ensure code-domain orthogonality between participating operators. The different networks in the coordination pool might need to exchange some of the CSI acquired from the UEs. In general, there is a tradeoff between the CSI accuracy and the required level of coordination, as a less accurate CSI requires a tighter coordination.

- Distributed synchronization schemes that help synchronize the BSs of participating operators within a geographical area. Synchronization of the networks participating in pooling helps avoid BS-BS and UE-UE interference that will be severe in deployments in which BSs of multiple operators are close to one another. Although some of this interference can be mitigated by directional transmissions, cellular broadcast channels and cell-wise reference signals can be severely hit by inter-operator interference, unless a sufficient level of synchronization among participating networks is ensured.

In Table II, we summarize the characteristics of the different architectural solutions and link them to the type of coordination, time-resolution, required supporting functions and information exchange overhead.

Table II: Summary of the characteristics of the different architectural solutions.

|  | Type of coordination | Time-resolution | Supporting function required | Information exchange overhead |
| --- | --- | --- | --- | --- |
| Interface at the RAN | distributed | real-time | enhanced CSI, distributed synchronization | high |
| Interface at the CN | distributed | long-term | enhanced CSI | low |
| RAN sharing | centralized | real-time | enhanced CSI | - |
| CN sharing | centralized | long-term | enhanced CSI | - |
| Spectrum broker | centralized | long-term | enhanced CSI | low |
| Uncoordinated | uncoordinated | - | enhanced CSI, spectrum sensing and DFS/DCS | - |

## 5 Open Issues

### 5.1 Beyond exclusive and pooled access: the emergence of hybrid spectrum regimes

In Sections 3 and 4, we focused the discussion on spectrum pooling. However, other spectrum sharing regimes are possible and should be further explored, see Fig. 4.

Under a license-exemption regime, there is no limitation to the number of operators sharing the spectrum. In other words, for the scenarios considered in this regime, the main difference between spectrum pooling and license-exempt is that in the first case the fixed number of operators sharing the spectrum allows a tighter control of the inter-operator interference.

There are recent technologies that aggregate carriers at both licensed and license-exempt spectrum, in a way to route the different information pipes to the carrier that best matches their requirements (for example, licensed assisted access (LAA) and LTE/WiFi link aggregation (LWA)). We refer to this as *hybrid spectrum regime*. A similar approach, based on a hybrid use of pooling (or license exempt) at mmWave and exclusive spectrum allocation at traditional cellular frequencies could also be exploited, see Fig. 4. For example, carrier aggregation could be used to transmit more critical information (for example, control and synchronization signals [9]) over the licensed spectrum, while sending less critical information over pooled spectrum at mmWave. We note that more work is needed to compare the different options.

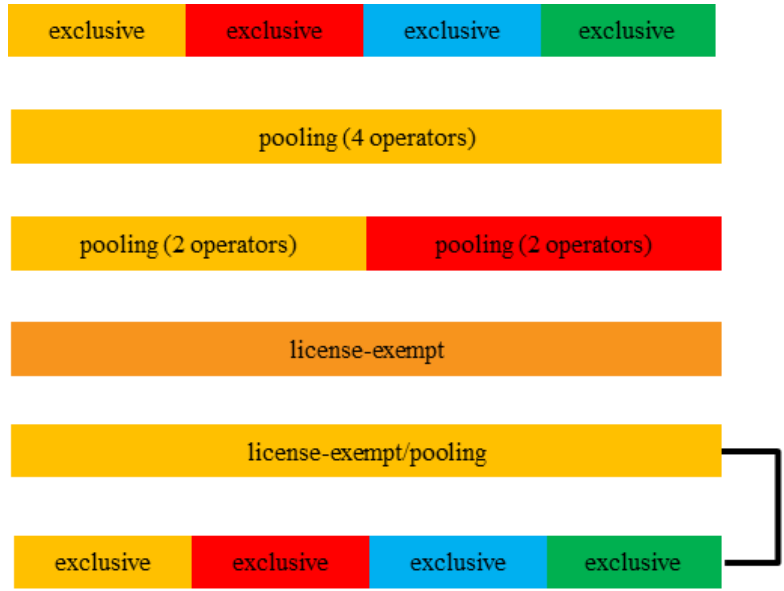

Fig. 4. Examples of spectrum sharing regimes.

### 5.2   Impact of real-world factors

The results given in Section 3 have been obtained under ideal assumptions. There is clearly a need for further work to better understand the impact of real-world factors, including imperfect CSI (that critically affects the beamforming accuracy), realistic antennas, backhaul latency, BS synchronization, and distributed coordination.

### 5.3   Impact of non-technical factors

The choice of the spectrum authorization type does not only depend on technology factors, which are the focus of this paper. Other factors include, for example, the desirability of promoting competition, encouraging investments and innovation, and achieving a widespread availability of services across rural and urban areas. Further work is needed to further study these and other non-technical factors.

# 6  Conclusions

MmWave communications have recently emerged as a solution to the spectrum scarcity in bands traditionally used for cellular communications. However, even at mmWave frequencies, the spectrum is not unlimited, which means it is essential to achieve an efficient use of the spectrum. In this paper we have shown that spectrum pooling at mmWave could allow a more efficient use of the spectrum than a traditional regime where exclusive spectrum is allocated to individual operators. In particular, we assessed the benefit of coordination among the networks of different operators, studied the impact of beamforming both at the base stations and at the user terminals, and analyzed the pooling performance at different frequency carriers. We have also discussed the technical enablers that are required to make spectrum pooling work under realistic assumptions and constraints, including the type of supporting architecture, the type of coordination, the amount and type of information exchange required, along with new functionalities.